# The Virtual Doctor: An Interactive Artificial Intelligence based on Deep Learning for Non-Invasive Prediction of Diabetes

Sebastian Spänig[1], Agnes Emberger-Klein[2], Jan-Peter Sowa[3], Ali Canbay[3], Klaus Menrad[2], Dominik Heider[1,*]

1: Department of Bioinformatics, Faculty of Mathematics and Computer Science, Philipps-University of Marburg, Hans-Meerwein-Str. 6, 35037 Marburg
2: Chair of Marketing and Management of Biogenic Resources, Weihenstephan-Triesdorf University of Applied Sciences/TUM Campus Straubing for Biotechnology and Sustainability, Petersgasse 18, 94315 Straubing
3: Department of Gastroenterology, Hepatology and Infectious Diseases, Otto-von-Guericke University Magdeburg, Leipziger Str. 44, 39120 Magdeburg

*corresponding author: dominik.heider@uni-marburg.de

## Abstract

Artificial intelligence (AI) will pave the way to a new era in medicine. However, currently available AI systems do not interact with a patient, e.g., for anamnesis, and thus are only used by the physicians for predictions in diagnosis or prognosis. However, these systems are widely used, e.g., in diabetes or cancer prediction.
In the current study, we developed an AI that is able to interact with a patient (virtual doctor) by using a speech recognition and speech synthesis system and thus can autonomously interact with the patient, which is particularly important for, e.g., rural areas, where the availability of primary medical care is strongly limited by low population densities. As a proof-of-concept, the system is able to predict type 2 diabetes mellitus (T2DM) based on non-invasive sensors and deep neural networks. Moreover, the system provides an easy-to-interpret probability estimation for T2DM for a given patient. Besides the development of the AI, we further analyzed the acceptance of young people for AI in healthcare to estimate the impact of such system in the future.

# Introduction

Millions of people are treated in emergency rooms in hospitals every year. However, a significant proportion of the patients are non-emergency cases, forcing the hospitals to allocate medical personnel where it is not necessarily needed, creating a suboptimal use of staff and treatment of real patient emergencies (Ismail *et al.*, 2013; Eastwood *et al.*, 2017; O'Keeffe *et al.*, 2018). Moreover, there is a growing shortcoming of physicians in rural areas, leading to underserved patients, in particular, due to the demographic change and a growing number of elderly people (Thomas *et al.*, 2015; Eidson-Ton *et al.*, 2016; Deligiannidis, 2017). One solution to overcome these problems in the future might be the use of artificial intelligence (AI) in healthcare as a driver for systems medicine (Baumbach and Schmidt, 2018). AI as speech recognition systems, such as Apple's Siri or Amazon's Alexa, have already entered our daily lives. They can be used as personal assistants or as an interface to interact and control smart homes. Moreover, speech recognition systems are also employed to support physicians in their daily routine in hospitals, e.g., to provide a computer-assisted documentation, i.e., they use speech recognition to translate speech to text and dictating patients' anamnesis results to fill in electronic health records. However, so far, speech recognition has not found its way into clinical decision-support systems. Today, clinical decision-support systems that are frequently used in clinical practice are rare, as they require expert knowledge and most of them have not been comprehensively clinically evaluated. However, there are some examples, including image analysis software for, e.g., magnetic resonance or computed tomography images (Bibault *et al.*, 2016), or expert systems that are used to predict treatment options, e.g., in infectious diseases, which are fully integrated into daily clinical use (Lengauer and Sing, 2006). The clinical use of decision-support systems is still awaiting a significant paradigm shift characterized by a virtuous cycle of systematic links and mutual feedback between "big data" acquisition, modeling, and acceptance. This sought-after revolution will come from a transdisciplinary effort involving the sharing of common representations and the design of big data acquisition and management, model developing and testing methods.

AI can improve medical treatment or diagnosis, e.g., HIV drug resistance prediction (Riemenschneider *et al.*, 2016), breast cancer prediction (Montazeri *et al.*, 2016), or type 2 diabetes mellitus (T2DM) (Talaei-Khoei and Wilson, 2018), however, these models cannot be used in an automated and unsupervised manner. For instance, prediction models for T2DM have been studied widely, e.g., Perveen *et al.* (2016) evaluated the performance of T2DM prediction based on ensemble learners. They employed data obtained from the Canadian Primary Care Sentinel Surveillance Network (http://cpcssn.ca), which includes demographic information, body mass index (BMI), high-density lipoprotein (HDL), and triglycerides. Anderson *et al.* (2016) analyzed different risk factors for developing T2DM in order to develop a biomarker panel. The final panel included blood sugar, age, race, triglycerides, BMI, and blood pressure. Kälsch *et al.* (2015) investigated possible correlations of liver enzymes and T2DM. The authors showed the predictive capability of logistic regression models based on HbA1c and adiponectin or HbA1c and BMI. Chen and Pan (2018) employed boosting algorithms to predict T2DM. In contrast, Läll *et al.* (2017) analyzed genetic associations in a genome-wide association study to identify risk factors for diabetes. In another recent study, data from 14,000 smartwatch users has been analyzed and it has been demonstrated that diabetes can be predicted by using the ability of the smartwatch to perform a cardiogram. These findings are mainly based on the eHeart study by Tison *et al.* (2018). However, most available AI systems are used by a physician in order to analyze patient data and to make predictions on possible outcome of a treatment or to provide additional data for a diagnosis or prognosis. These models are typically based on statistical and machine learning approaches, however, they are not coupled with speech recognition or synthesis to interact with a patient. The aims of the current study were 1) the development of an AI able to interact with a patient (virtual doctor), and 2) to demonstrate its usefulness on the automated prediction on an example disease, namely T2DM, on a large patient cohort, and 3) the design and analyses of a questionnaire on the acceptance of AI in medicine by young adults. The development of an AI is also faced with the problem of the uncanny valley (Mori *et al.*, 2012), i.e., AI mimicking human behavior elicit eeriness and revulsion among some observers. However, there is already some research that suggests these problems might be generational, as young people more used to AI may be less likely to be affected (Hanson *et al.*, 2005). Thus, we focused on young adults in the current study to analyze acceptance and intention to use AI in medicine.

## Materials and Methods

### Design of the Virtual Doctor

**Overall Concept:** The aim of the study is the automated, interactive anamnesis with a non-invasive T2DM prediction as a proof-of-concept. The virtual doctor is a cabin with several devices to obtain patient metrics. The embedded AI utilizes these metrics, such as the patient's BMI, to identify possible diseases or impairments to health. Finally, the AI recommends further diagnostic steps to the medical personnel, such as the HbA1c blood test in our example study. Currently, besides the measurement of the patient's weight and height, a speech recognition facilitated dialog system, interviews the patient to determine the patient's age and gender. Furthermore, additional anamnesis questions about the alcohol and tobacco consumption allows the AI to incorporate further T2DM risk factors. The hardware is assembled to a patient cube, which serves as a frame for the respective measurement devices. The prototype is designed as a self-contained system which allows an easy extension for supplementary devices. Within the anamnesis, the obtained information will be submitted and evaluated on a computer.

**Detection of Weight and Height:** The weight $w$ is gained through a scale which has been adapted in a way so that data is submitted directly to the computer (Figure 1). In order to calculate the body height $h$, the distance $d$ from the ceiling to the patient's head is subtracted from the total height (200 cm). The distance $d$ is recorded by an ultrasonic sensor (Figure 2):

$$h(d) = \frac{200-d}{100} \tag{1}$$

Afterwards, the BMI is calculated by the following formula:

$$BMI(w, h) = w/h^2 \tag{2}$$

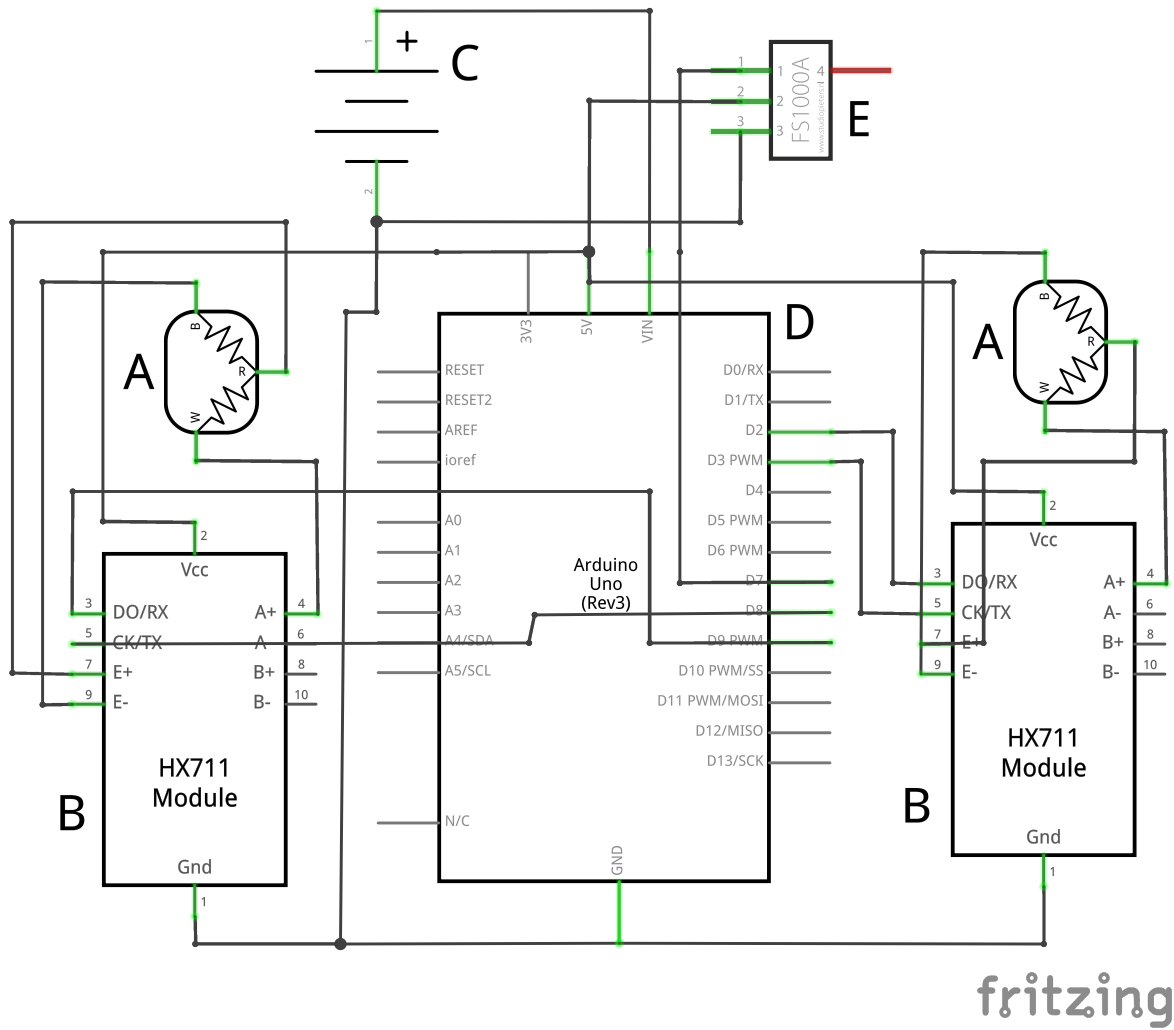

*Figure 1*: The image above shows the circuit diagram of the scale. The main components are two load cells, each capable to measure up to 200 kg (A). Two HX711 sensors (B) convert the analogue signal from the load into digital signals and forward them to the Arduino (D). A wireless RF Link Transmitter broadcasts the weight to the server (E). An external power source is necessary to supply the devices with electricity (C).

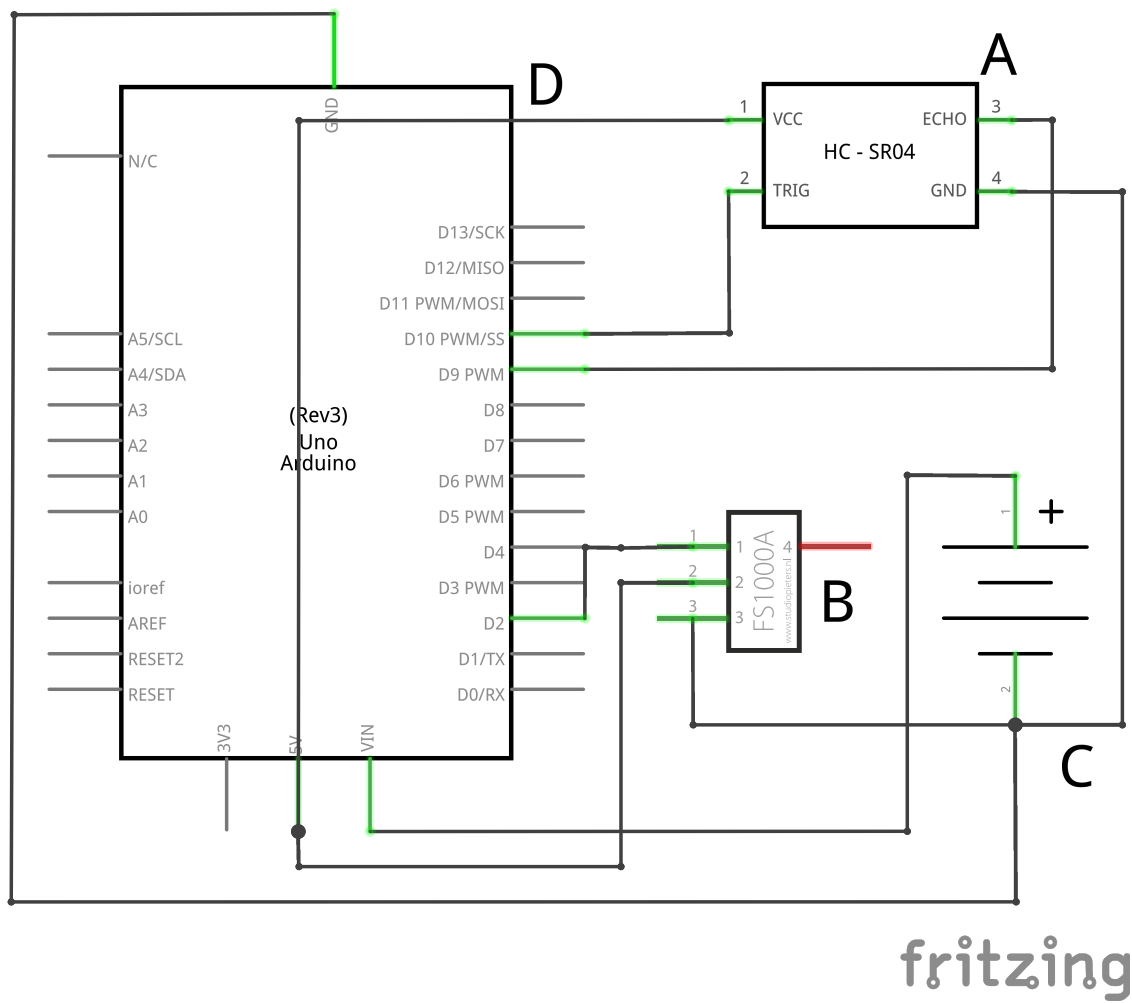

*Figure 2*: This draft shows the circuit diagram of the height meter. The core component is the HC-SR04 ultrasonic sensor (A). It triggers a ultrasonic wave, which is reflected by an object. A further sensor receives the returning signal and the time in between is transmitted to the Arduino (D). The distance between the sensor and the object can be calculated by means of the duration and finally forwarded via the RF Link Transmitter to the server. A 9 V battery is again responsible for the power supply (C).

**Speech Recognition:** We used CMUSphinx for speech recognition in the current study (http://cmusphinx.github.io). In order to facilitate speech recognition, the necessary data for the German language has been obtained. That includes the German language model, the acoustic model, and a dictionary from an open-source repository, which collects transcribed speech for application in speech recognition (VoxForge). Additionally, CMUSphinx´ Pocketsphinx as well as Sphinxbase are employed to transform the patient answers to their respective string representation.

The speech signal is conveyed as a pre-recorded file. Therefore the patient answers are recorded before the actual recognition takes place. It is necessary to adjust the record settings to the determined conditions in the available acoustic model. In particular the recording is conducted through one channel (mono) on a frequency of 16 kHz and finally a bit depth of 16, which controls the resolution of the record.

It is imperative to obtain correct answers and to minimize repetition of questions in case of an incorrect recognition. Furthermore, the virtual doctor's location is associated with certain properties in terms of the acoustic, especially background noises or the way the walls echo speech signals. Moreover, the used microphone alters the recordings and the quality of the obtained speech varies due to distance variations between the patient and the microphone. To overcome such difficulties, we adapted the default acoustic model based on the tutorial provided by the CMUSphinx developers.

**Diabetes Data:** For the development of the T2DM prediction model, we used the data previously analyzed by our group (Kälsch *et al.*, 2015). The data was collected during the Heinz-Nixdorf-Recall study (HNR) (Schmermund *et al.*, 2002). The HNR cohort is population based and includes 4814 participants (2419 female), aged 45 to 75 years (female 59.6 ± 7.8y, male 59.7 ± 7.8y). Men exhibited significantly higher weight (86.2 ± 13.2 kg; p < 0.0001) and waist circumference (100.3 ± 10.8 cm; p < 0.0001) than women. The mean BMI of the cohort was 27.9 kg/m², suggesting large parts of this population to be overweight or obese. Both women and men were considered overweight by BMI, with men reaching higher values (male: 28.2 ± 4.0 kg/m² vs. female: 27.7 ± 5.2 kg/m²; p < 0.0001). Detailed demographics are given in Table 1.

*Table 1: Demographics of the HNR cohort*
*(\*\*\* p < 0.001)*

|  | male (n=2395) | female (n=2419) |
|---|---|---|
| age (years) | 59.7 ± 7.8 | 59.6 ± 7.8 |
| height (cm) | 174.8 ± 6.8 | 162.1 ± 6.2*** |
| weight (kg) | 86.2 ± 13.2 | 72.6 ± 13.8*** |
| BMI (kg/m²) | 28.2 ± 4.0 | 27.7 ± 5.2*** |
| waist circumference (cm) | 100.3 ± 10.8 | 88.5 ± 12.9*** |
| Diabetes n (%) | 418 (17.5) | 238 (9.8)*** |

T2DM was present in 656 individuals of the HNR cohort. Thereof 397 (8.2%) had previously known T2DM and 259 (5.4%) had unknown diabetes. T2DM was more common in men with 418 (17.5%) male subjects affected compared to 238 female subjects (9.8%). The highest proportion of T2DM was found in subjects with BMI above 40 (approx. 45%).

**Machine Learning:** In our study we employed deep neural networks (DNNs) and support-vector machines (SVMs) for the prediction of T2DM. To this end, we used the R packages *deepnet* and *kernlab,* respectively. To reduce the bias due to the class imbalance in the dataset, we used a sub-sampling approach (Fithian and Hastie, 2014). We separated the data into training and test data (80:20). Each training instance is represented as $\langle \vec{x_i}, y_i \rangle$ with $\vec{x_i} \in (sex, age, bmi)$ and $y_i \in \{0,1\}$ or with $\vec{x_i} \in (sex, age, bmi, HbA1c)$ and $y_i \in \{0,1\}$, respectively. Before training of the networks, we normalized the data by using a z-transformation:

$$z = \frac{x - \acute{x}}{\sigma} \qquad (3)$$

We tested DNNs ranging from one up to twenty neurons in the hidden layers. Moreover, we analyzed the performance of the DNNs with one up to three hidden layers. We used the *tangens hyperbolicus* as activation function, whereas a sigmoid output function was used. The DNNs were trained using 100 training epochs. For the SVMs we used an RBF kernel with default parameters.

**Statistical Analyses:** For evaluation of the models, we calculated the Receiver Operating Characteristics (ROC) and the corresponding Area Under the Curve (AUC). For comparisons between the different models we used the method of DeLong *et al.* (1988) using the R package *pROC* (Robin *et al.*, 2011).

We used an one-sided Student's t-Test on the AUC distributions to test whether a model outperforms simple guessing. Thus, the alternative hypothesis states that the mean of the AUC from our model is greater than 0.5 and thus better than assigning classes by chance. Furthermore, the models were tested for significance using a permutation test. For a permutation test, the class labels have been randomly permuted and the AUCs have been calculated (Sowa *et al.*, 2014).

**Probability Estimations:** With machine learning (ML) models becoming more complex and their results less comprehensible even for ML-experts, there is the unmet need for verifiable algorithm output. Many ML models are perceived as black boxes by non-experts and, unfortunately, the obtained classifier scores cannot usually be directly interpreted as class probability estimates. In probability-focused medical applications, it is not sufficient for a classifier to perform well with regards to class discrimination and, consequently, we employed the calibration method GUESS, implemented in the R package *CalibratR*, in order to enable probabilistic interpretation (Schwarz and Heider, 2018).

Calibration of a set of ML scores $X = \{x_1, \ldots, x_n\}$ is defined as calculating $p_{cal}$ for $x_i$. The aim of calibration is to find a value $p_{cal_i}$ for the original input $x_i$ that satisfies the equation

$$p_{cal_i} = p(y = 1 \vee x_i) \qquad (4)$$

Based on above notation, a well-calibrated prediction can be defined as follows: A probability estimation $x_i$ is well-calibrated if the event $y_i = 1$ actually happens with an observed relative frequency consistent with the prediction $x_i$. Consequently, a well-calibrated ML model outputs probability estimates $x_i$ that are consistent to the conditional probability $p(y = c \vee x_i)$, the long-time empirical frequency for observation $x_i$ to be of class $y = c$. GUESS uses a probability density fitting algorithm based on highest log-likelihood and returns calibrated prediction, i.e., probabilities. We used GUESS in order to estimate probabilities based on the machine learning models mentioned before.

The probability estimates of the model based on sex, age, and BMI are then modified based on the anamnesis questions. In the current version we use the following anamnesis questions:
1. Do you suffer from an increased desire to void your bladder?
2. Do you suffer from an extraordinary increased thirst?
3. How would you rate your alcohol consumption behavior? (1 to 10, 1: not present/very little, 10: a lot)
4. How would you rate your tobacco consumption behavior (1 to 10, 1: not present/very little, 10: a lot)

For tobacco and alcohol consumption, the probability for class 1 (i.e., T2DM) is increased with increasing usage. For polyurea and polydipsia, the probability is either increased or decreased depending on a positive or negative answer. If the final probability lies within a twilight zone, i.e., between 30 and 70 %, the virtual doctor recommends the HbA1c blood test in order to confirm or reject the initial diagnosis.

## Questionnaire design

**Study, Data Collection and Privacy Statement:** We conducted a prospective cross-sectional self-administered paper-and-pencil survey using a standardized questionnaire. To collect the data students from three different German universities (Technical University of Munich (TUM), University of Applied Science Weihenstephan-Triesdorf (HSWT), and University of Marburg (UMR)) and different subjects of study (e.g., food sciences, horticultural sciences, computer science, and engineering) were asked to answer the paper-and-pencil questionnaire in class from November 2017 to January 2018. Before answering the questionnaire, all interviewees were informed about the aim of the study and the privacy policy. All participants gave informed consent, the participance was voluntarily. The data privacy statement and the documentation of the procedure (procedure index) were approved by the Data Protection Officer of the HSWT before the survey. Ethics approval was not applicable for the type of human data presented in this study. We followed the ethical principles of the Helsinki Declaration, the German Research Foundation (DFG), and the German Society for Sociology (DGS).

**Questionnaire Design, Sections and Items:** The questionnaire started with general questions about AI (e.g., awareness of the term, attitudes towards and usage of AI). Section 2 contained questions about the usage of voice assistants, while section 3 focused on the topic *AI in the health sector/in medicine*. This vital section included questions about the usage intention of AI in medicine, about the ease of use and perceived usefulness of this technology, and about attitudes toward AI in medicine. Additionally, participants had to specify procedures, which they could imagine receiving from AI as well advantages, disadvantages, and necessary precondition of the use of AI in medicine. To measure usage intention, we followed the study of PwC (2017) and asked the students "to consider advanced computer technology or robots with AI that had the ability to answer health questions, perform tests, make a diagnosis based on those test and symptoms, and recommend and administer treatment" (PwC, 2017). Subsequent interviewees had to answer on a 5-point scale form *I would certainly engage with* to *I would certainly not engage with* how willing they were to engage with this technology in future.
At the end of the survey participants had to name in section 4 their subject of study and had to answer questions about socio-demographic data, such as gender and age. The design of the questionnaire was critically reviewed before the final survey using a paper-and-pencil pre-test to guarantee clarity and comprehensibility of the instrument.

**Data and Data Cleaning:** Originally, n=330 individuals participated in the survey. However, ten questionnaires were excluded due to eight or more unanswered (sub-)questions. To guarantee full datasets, missing values of ordinal variables were imputed with the median of the variable. Additionally, in the question dealing with the pre-requisites of using AI, some individuals (n=23, question 3.7a or n=25, question 3.7b ) marked more than one answer-category. In this case the answer which describes the shortest waiting time was used. After data cleaning, n=320 cases could be included in the analyses. The questionnaire and anonymized data of the presented results can be found as supplement 1 and 2, respectively.

## Results

### Diabetes Prediction Model
Figure 3 shows the overall workflow for the prediction of T2DM. Gender and age are acquired via speech recognition. Therefore, the CMUSphinx toolkit has been employed to allow reliable prediction of numbers between 1 and 100 as well as the categories male, female, yes and no. Whereas the numbers from 1 to 10 also serve as a severity index, from 1 – not present/very weak towards 10 very strong. The trained model will be fed with the anamnesis results and returns the probability that a patient may suffer from a certain disease, which is in the present case type 2 diabetes mellitus. If the result exceeds a certain threshold, i.e., is within a twilight zone, the virtual doctor suggests the HbA1c blood test in order to confirm or dismiss the diagnosis (Figure 3).

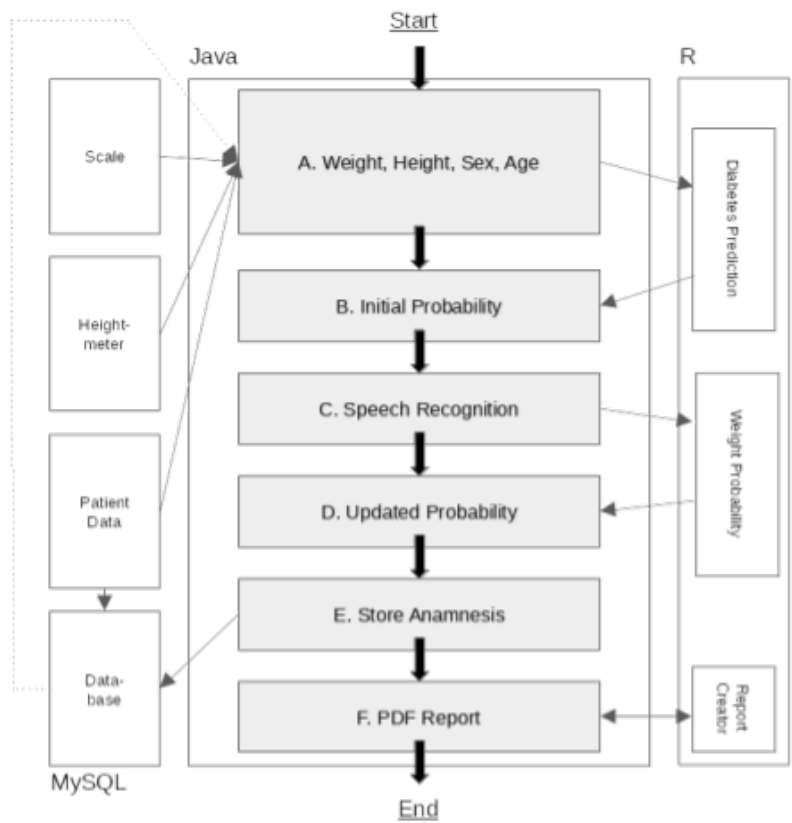

*Figure 3*: *The figure above illustrates the workflow from the beginning of the anamnesis until the end. The respective rectangles are annotated with the corresponding implementation domain. In future versions the patient history might be additionally incorporated (dotted line to A).*

The DNN trained without HbA1c outperforms the SVM in terms of AUC (0.703 vs. 0.679, p<0.05). When using different number of neurons in the hidden layer in the DNNs, the AUC quickly converged as shown in Figure 4. Adding more than one hidden layer did not significantly improve prediction performance in terms of AUC.

We used GUESS in order to provide calibrated prediction that can be interpreted as probabilities. It turns out that GUESS could slightly improve the calibration for the DNN from an ECE of 0.07 to 0.05. However, the calibration was not improved for the SVM, the SVM alone as well as combined with GUESS reached an ECE of 0.06. This is mainly due to the fact that the *kernlab* package already provides a probability estimation for SVMs, which is based on plat scaling (Lin *et al*., 2007). However, the ECE of the DNN is smaller compared to the ECE of the SVM while having a better discriminative power in terms of AUC. Thus, we selected the DNN instead of the SVM for the initial prediction of T2DM.

DNN and SVM exhibit similar prediction performance (AUC=0.84 and AUC=0.825, respectively; p=0.299), when trained with HbA1c. This is in line with existing models, in particular with the model by Kälsch *et al*. (2015), which was trained on the same dataset (AUC=0.851). The models were calibrated using GUESS, however, it turned out that GUESS was not able to further improve the models in terms of calibration.

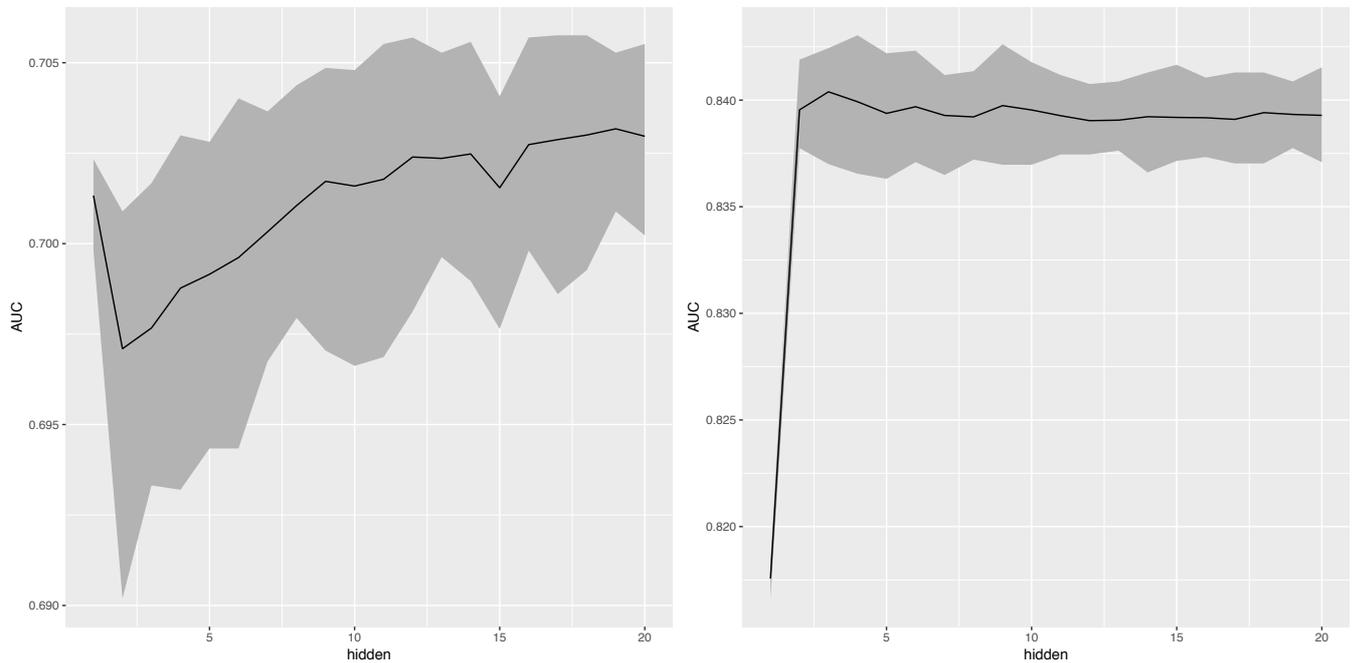

*Figure 4*: Convergence of the DNN with one hidden layer and different number of hidden neurons. Left: DNN without HbA1c; right: DNN with HbA1c.

## Young Adults' View of AI in Medicine

Table 2 shows important socio-demographic (SOD) characteristics of the sample as well as information about the subject of study. Approximately two thirds of the respondents were male and 85% were younger than 25 years old. Most of the students studied Engineering Science (36.6%) or Agronomy / Forestry / Nutritional science (37.5%). Compared with the total population of students in Germany there existed an overrepresentation of male students and students studying Agronomy / Forestry / Nutritional sciences in the sample.

*Table 2: Socio-demographics of young adults*

| n=320 | | | | | |
|---|---|---|---|---|---|
| **Gender (%)** | female | 32.8 | **Field of study (%)** | humanities, law, economics, social sciences | 8.4 |
| | male | 67.2 | | agronomy, forestry, nutritional sciences | 37.5 |
| **Age (%)** | 18-20y | 49.7 | | engineering, computer science | 36.6 |
| | 21-<25y | 35.3 | | mathematics, natural sciences | 15.9 |
| | 25-<30y | 13.1 | | others/not known | 1.6 |
| | 30+y | 1.9 | | | |

Most of the respondents stated that they have heard the term AI before and were able to explain what it means (56.3%), while further 40.9% stated that they were not able to explain exactly what it means. Additionally, more than half of the respondents had a (very) positive attitude towards AI in general. In contrast, only a minor part had already used some form of AI, e.g., a cleaning robot (2.5%) or a health assistant (5%) in the past. However, in the case of the latter further 41.6% stated that they would like to use this form of AI in future.

Concerning the health context it is interesting that some of the participants (19.7%) already got in touch with a fictitious artificial doctor, since they knew the Emergency Medical Hologram (EMH) from Star Trek Voyager. Thereby, awareness was higher for male compared to female students (Chi²=12.213; p<0.001). Of those students, who were aware of the EMH more than half of the respondents (53.7%) would not hesitate to receive treatment from him. When asked for the intention to use an AI for healthcare needs the participants showed on average a slight positive to medium willingness to engage with this kind of technology in future (mean: 2.6, sd: 1.17) (Table 3). However, intention varied between socio-demographic groups: Male respondents showed a higher intention than female (mean$_{male}$: 2.4 vs. mean$_{female}$: 3.0; t=4.811, p<0.001). There was also a tendency that younger students had a higher intention compared to older students (mean$_{18-24y}$: 2.5 vs mean$_{25+y}$: 2.9; t=-1.865; p<0.1). This is in line with the finding that the uncanny valley (Mori *et al.*, 2012) may be linked to age (Hanson *et al.*, 2005).

Additionally, intention differed by field of study (F=6.744; p<0.001): The highest intention was found for students studying engineering / computer science (mean: 2.3) or mathematics / natural science, while the lowest intention was observed for students studying agronomy / forestry / nutritional science (mean: 2.9). Post-Hoc-Tests (GT2 following Hochberg) showed that intention significantly differed between Engineering Science students and students from the field of Agronomy / Forestry / Nutritional sciences (p<0.001) as well as between those students and Mathematics / Natural Science students (p<0.05).

*Table 3: Description of the variable Intention by socio-demographic and field of study data*
*m mean, sd standard deviation; \* p < 0.05; \*\* p < 0.01; \*\*\* p < 0.001; $^a$ Please imagine some advanced computer technology or robot with AI which has the ability to answer health questions, perform tests, make a diagnosis based on those test and symptoms, and recommend and administer treatment. If you had the opportunity to make use of this technology, how willing would you be to engage with it in future? $^b$Independent group t-test; $^c$One-way ANOVA (individuals with unclear field of study were excluded, n=316); $^d$I certainly would make use of it (=1) to I certainly would not make use of it (=5).*

| n=320 | | mean | Sd |
|---|---|---|---|
| **Intention$^a$** | total sample | 2.6 | 1.17 |
| **Sex$^b$** | Female | 3.0*** | 1.14 |
| | Male | 2.4*** | 1.13 |
| **Age$^b$** | 18-24y | 2.5 | 1.18 |
| | 25+y | 2.9 | 1.07 |
| **Field of study$^c$** | humanities, law, economics, social sciences | 2.5*** | 1.07 |
| | agronomy, forestry, nutritional sciences | 2.9*** | 1.21 |
| | engineering, computer science | 2.3*** | 1.07 |
| | mathematics, natural sciences | 2.4*** | 1.15 |

# Discussion

The current situation of medical care in industrialized countries is characterized by unmet demands in several fields. As described above, utilization of emergency services in actual non-emergency cases is common. This puts medical personnel and resources under high pressure and costs time, which could be allocated to true emergencies. Another problem arises in rural areas, where the availability of primary medical care is strongly limited by low population densities. On the one hand this requires long distance trips to medical professionals for the patients and on the other hand makes it unattractive for medical professionals to provide services in such an area due to very ineffective allocation of time and resources. A third critical factor in medical care is a very high workload for general practitioners in urban areas. Due to the high density of population, most larger cities have low general practitioners to population ratios. This results in a high number of patients, which have to be seen by each practitioner per day, lowering time and resources available for each patient. These three problems of current medical care could all be ameliorated by a partially autonomous AI for anamnesis and initial medical diagnosis. In particular in high pressure settings, as in emergency settings and in highly frequented general practitioners, the time-consuming process of anamnesis could be automated and thus standardized. Non-emergency cases could be identified early and treated as such, i.e., by recommending to see a practitioner or specialist during normal operating hours. This would free much needed time for the medical personnel. In rural areas an autonomous medical AI could serve as an initial point of care for patients with supposed need for medical assistance. The AI could either point the patient to the closest emergency service, specialized for the situation, or advise to see a practitioner in a non-emergency setting or even provide assurance that no medical condition is given that would require attention of a medical doctor. By automation of the basic process required in all types of medical care, anamnesis, initial assessment of situation and diagnosis, it would be possible to provide such a service in a very standardized fashion available at any time.

In the present study we demonstrate the general feasibility to automate certain diagnostic procedures by an AI, in this case of the presence of T2DM. The DNN-based AI was able to detect presence of T2DM with an AUC comparable to the state-of-the-art based only on demographic and basic physiological (non-invasive) data, which can be easily and automatically measured with external sensors coupled with speech recognition.

For a small fraction of patients, this diagnosis could be seen as definite without need of further tests as HbA1c or an oral glucose tolerance test, which would always be performed by a human medical professional to ascertain suspected T2DM during anamnesis. In this sense the virtual doctor already outperforms human medical professionals for this specific disease. In any case the system can be assistant or even proxy for a medical professional during the process of anamnesis, suggesting an initial diagnosis and recommending further diagnostics to assure the assessment. By further optimizing the underlying AI, a fully assistant function of the system for a medical doctor could be achieved, that may recommend multiple additional diagnostic steps and even treatment.

We could further show that DNNs performed better than SVMs for the prediction of T2DM when trained without HbA1c and equally when trained with HbA1c. This might not be the case for all settings and diseases and has to be explored for every additional diagnostic step or possible disease state, that should be detectable by the system. Moreover, we were able to provide reliable probability estimations for the DNN predictions, which improves interpretability of the DNNs for the physicians as well as for the patients.

AI is able to assist physicians in the diagnosis of diseases or therapy recommendations. However, existing tools do not provide an interactive diagnosis for the patient. Interaction of the AI with the patient comes with the risk of the uncanny valley (Mori *et al.*, 2012), i.e., AI mimicking human behavior elicit eeriness and revulsion among some observers. It has been shown in some studies that these impressions are generational and that younger people, who are more used to AI, may less likely to be affected (Hanson *et al.*, 2005). This seems to be true also in our study, were more than half of the analyzed individuals have a positive connotation with AI and are open for AI in medicine. This can be particularly observed when comparing the intention to use AI in healthcare between the groups aged 18-24y and 25+y (2.5±1.18 vs. 2.9±1.07). However, there seems to be a trend that students from natural science / mathematics and engineering / computer science have a higher tendency compared to the other study subject groups. These differences could also be observed for gender. Males do have a higher tendency compared to females (2.4±1.13 vs. 3.0±1.14) as well as awareness of the Emergency Medical Hologram (EMH) from Star Trek Voyager ($p<0.001$). However, gender could also be a co-founding factor regarding the tendency to use AI and the study subject.

AI can contribute widely to medicine in the future, e.g., by providing first-line diagnosis in hospitals, but also to counteract shortage of physicians in rural regions for non-emergency patients. In the latter case, AI could provide diagnosis of non-emergency diseases, access to drugs only available by prescription, and could provide a doctor's note for the employer. In the future we will employ other types of sensors, e.g., infrared cameras, to improve the prediction results and to broaden the spectrum of diseases, e.g., for the automated detection of acute infections.

The presented prototype system of a medical AI has several features that have the potential to significantly improve overall medical care. One possibility would be to integrate additional non-invasive diagnostics. Measurements of blood pressure, blood glucose levels, and heart rate can be measured in an automated manner and would be an important addition to a fully autonomous medical diagnosis system. It might even be possible to read out data from health monitors or smart watches with respective functions already in use by many people. Detectors which are still somewhat experimental, i.e., an electronic nose for detection of components in breath to identify possible lung diseases, tumors, etc., could be integrated in later iterations of the system. Another example for such new methods is the diagnosis of anemia via estimation of hemoglobin levels from photos of finger nail beds (Mannino *et al.*, 2018). Simple, visually based diagnostic systems thus would be easy to integrate into an autonomous medical AI. In addition, the speech recognition system and also the speech output are not limited to a single language. In fact, given sufficient storage and computing capacity an autonomous medical AI could speak any language a patient may speak and thus overcome language barriers that may arise between patient and doctor. The result of anamnesis, initial diagnosis, and recommendation for further procedure can easily be printed in the mother-tongue of the patient and of the medical professional in parallel. Finally, an autonomous medical AI with advanced non-invasive diagnostics can be positioned at any desired location, i.e., in rural areas. Such a "standalone" medical AI could provide basic medical care in areas of low density for medical professionals and not only point patients to relevant and needed expert care but in parallel alert emergency facilities in cases of true urgent medical emergencies.

There are still several limitations to the idea of an autonomous medical AI. The most important limitation is that such a system will depend strongly on acceptance and consent by patients and also by medical professionals. We present survey data from young academics that a relevant proportion of these would use such a system, if they had an unclear medical condition. This has to be expanded for other populations, in particular for individuals with lower education status and higher age. In particular, we will collect data on possible acceptance of such a system by medical professionals in a follow-up study. If these would not trust the assessment or the recommendations of a medical AI, the reduction in time and resource allocation for an individual patient and the resulting benefit would be limited. Other important aspects, which have to be clarified for such a system, are matters of data storage, data transfer, and data integrity compliant with respective laws of data protection in countries willing to implement such a system. In settings were such a system is applied for emergency rooms, hospitals or general practitioners, this would be a minor problem, as the data can be directly transferred into existing systems, making use of patients consent to data storage and utilization for this specific medical service provider. An organizational problem might become the combination of multiple sub-AIs for a range of detection methods (e.g., the electronic nose), disease states, and storage for multi-language speech recognition and output. Computing and data storage capacities will have to reflect the complexity of the system as additional diagnostic tools will be connected. However, this might be compensated by a parallel progress of faster and smaller processors and data storage facilities.

In summary we present data on an autonomous medical AI, the virtual doctor, as an assistant for anamnesis and diagnosis of T2DM, able to recommend further diagnostics to a medical professional. Such a system would receive acceptance among young academic individuals, if they would be patients with an unclear medical situation. An AI as the presented one could ameliorate unnecessary utilization of emergency utilities, reduce workload of general practitioners, and could possibly improve the medical service in rural areas with low density of medical professionals. Integration of current and under development non-invasive detection methods would pave the way for Medicine 4.0. Given consent of the patients, the data acquired by this system in a highly standardized way would improve reliability of clinical studies and population-based cohort studies.

## Supplementary Materials
Additional file 1: Questionaire and coding of variables
Additional file 2: Dataset from survey

**Acknowledgments**: We thank the students for answering the questionnaire.
**Funding**: This work was supported in part through funding by the Deichmann Foundation (DH and AC) and by the Wilhelm-Laupitz-Foundation to AC.
**Author Contributions**: SS, AEK, KM, and DH designed the study, analyzed and interpreted the data. SS, AEK, and DH wrote the manuscript. JS and AC provided clinical information and analyzed and interpreted the data. All authors read and approved the final manuscript.

**Competing Interests**: All authors declare that they have no competing interest.

**Data and Materials Availability**: The survey data has been published with the paper.